\documentclass[prl,floatfix,showpacs,a4paper,preprint]{revtex4}
\usepackage[dvips]{graphicx}
\usepackage[dvips]{color}
\usepackage{epsfig}
\usepackage{subfigure}
\usepackage{amsmath}
\usepackage{amssymb}
\usepackage{times}

\newsavebox{\CAuthor}
\begin{lrbox}{\CAuthor}
\footnotesize\verb+harmanda@mpip-mainz.mpg.de+
\end{lrbox}

\newcommand{\caplist}[1]{}

\renewcommand{\caplist}[1]{#1}

\begin {document}

\title{Quantitative Study of Polymer Dynamics Through Hierarchical
Multi-scale Dynamic Simulations}

\author{Vagelis A. Harmandaris}
\author{Kurt Kremer}

\affiliation{Max-Planck-Institut f\"ur Polymerforschung,
             Ackermannweg 10,
             55128 Mainz,
             Germany}

\date{\today}

\begin{abstract}
The long time dynamics of polymeric materials has been extensively
studied in the past through various experimental techniques and
computer simulations. While computer simulations typically treat
generic, simplified models, experiments deal with specific
chemistries. In this letter we present a hierarchical approach
that combines atomistic and coarse-grained simulations to
\textit{quantitatively} study polymer dynamics. As an example we
predict diffusion coefficients of atactic polystyrene melts of
molecular weights relevant to polymer processing (up to 50kDa)
\textit{without} any adjustable parameter and compare the results
to experiment.
\end{abstract}

\pacs{36.20-y, 61.25.H-, 83.10Rs}

\maketitle

\textit{Introduction.}$-$ The dynamics of polymers is a prototypical
soft matter problem and has attracted considerable experimental
and theoretical attention for many years
\cite{doiedw,ferrybook,binderbook,mcleish1}. For the two limiting cases of
short, but already fully flexible polymer chains in a melt, and
very long, fully entangled chains well established (scaling)
theories, namely the Rouse and reptation theory, exist and have
been tested in detail by experiments and computer simulations of
highly idealized models
\cite{doiedw,kk_erice,harmanda1,baschnagel}. However
most of these approaches provide either a semi quantitative link
between theory and experiment in the sense that scaling relations
are verified, or quantitative information for short chains with
simple chemical structure, like PE. What is still missing is a
direct quantitative link between chemical structure and measurable
quantities like diffusion constants or viscosities, especially for
high molecular weight entangled systems. In addition the crossover
regime from Rouse like to reptation like behavior, which for
example for polystyrene (PS) covers a range of molecular weights
from about $M \approx $1kDa to $M \approx $100kDa, is still under
discussion. In this regime it is also not clear, to what extent
universal theories describe the properties and to what extent
chemical details surface strongly, which makes the applicability
of a general analytic theory virtually impossible.  This is a
situation, where well tailored computer simulation can be of
significant help.

On the microscopic level detailed atomistic molecular dynamics
(MD) simulations allow quantitative predictions of the dynamics of
simple polymers with rather low molecular weight
\cite{harmanda1}. However, due to the broad spectrum of
characteristic lengths and times involved in polymeric systems, it
is not feasible to apply them to systems of more complex chemical
structure or of high molecular weight.

On the mesoscopic level coarse-grained (CG) molecular dynamics and
Monte Carlo simulations have proven to be very efficient means to
study the dynamics of long, entangled simple model polymer systems
\cite{kk_erice,baschnagel}. In a similar way, structure-based CG
models have been employed to study the dynamics of specific
polymer melts. There groups of chemically connected atoms are
lumbed into 'superatoms'. Effective, coarse-grained interaction
potentials are obtained by averaging over microscopic details of
the underlying atomistic models. The direct link to the chemistry
is maintained through the choice of appropriate superatoms and the
resulting set of structurally defined CG bonded and non-bonded
temperature dependent effective potentials (more precisely free
energies) \cite{kk_erice,actapaper1,vagelis2}. By doing
that structural properties of polymeric systems are described
quite well. However CG simulations cannot be used for a direct
\textit{quantitative} study of dynamics because the intrinsic time scale
of the CG model
is not the same as that of
the underlying chemical system. One reason is the use of (softer)
effective CG potentials, which results in a significantly reduced
effective friction between the beads compared to the original
monomers. To overcome this limitation the ratio between the CG
time and chemically realistic time scale has to be derived from
either experiments or atomistic simulations
\cite{actapaper1,vagelis2}.
However a direct comparison of the resulting dynamics to experiment
and the extension to long, entangled chains are still missing.

In this letter we extend this approach and present a hierarchical
methodology that combines dynamic simulations, on different length
and time scales. This approach allows us to predict through CG
simulations, not only \textit{qualitatively}, but also
\textit{quantitatively}, the dynamics of polymer systems of
molecular weight relevant to polymer processing \textit{without}
any adjustable parameter. As an example the whole methodology is
applied in atactic polystyrene at T=463K, which has been
extensively studied in the past through experiments (see for
example \cite{ferrybook,lodge1} and references within and
simulations \cite{actapaper1,vagelis2}).
The chosen temperature is a typical process for PS.

\textit{Simulation Methodology.}$-$
In order to derive a time scaling factor $S$ between different
models we match the mean square displacements of the monomers in
amplitude \emph{and} slope. The latter is important, since the
motion characteristics of the different models coincides only
above a characteristic length scale. A three level ansatz is
needed in order to also capture the effect of the chain length
dependent melt density onto the scaling factor.

Detailed all atom simulations, using a model in which hydrogens and carbons are
treated explicitly \cite{plathe2}, have been performed for rather
short PS chains ($M$=1kDa, $T$=463K, only 10 repeat units) at the
experimental density of 0.925gr/cm$^{3}$. Only for such small chains
reliable data for the mean square displacement of the monomers can
be obtained with all-atom MD simulations.
To somewhat overcome this problem united atom (UA) models are
frequently employed; here we are using the TraPPE one \cite{trappe1}.
There hydrogens are lumped together with carbons, defining new
CH$_{x}$ types of united atoms. For UA simulations
it is usually assumed that the time scale does not in a detectable
way deviate from all atom simulations, because hydrogens are very
small and the "coarse-graining" of the UA models is of the order
of hydrogen-carbon bond which is only about 1 \AA. However this
assumption, which in many cases (for example UA models for PE, PB,
PI) works reasonably well, fails to work for the UA-TraPPE PS
model; i.e. it predicts for PS a much faster dynamics. This is
most probably, due to the too low dihedral barriers or due the
missing electrostatic interactions but does not affect
the overall conformations \cite{vagelis2}.
Various systems, with molecular weight from 1kDa to 10kDa
($T$=463K), have been studied by this method. Beyond that CG
models are used, which however do not reproduce the motion
patterns down to length scales of the order of one or two
monomers.

Thus we adopt the following strategy. First UA and CG simulations
are performed, for molecular weights between 1 and 10kDa
in order to obtain the time scaling factor
$S_{UA-CG}(M)$. Then the UA results are calibrated by  all atom
simulations, however for much shorter times only, resulting in an
additional scaling factor $S_{AA-UA}(M)$. With $S(M)=S_{AA-UA}(M)
\cdot S_{UA-CG}(M)[sec/\tau]$ we calibrate the time scale to
determine long chain polymer diffusion constants and compare these
to experiment.

The molecular dynamics package GROMACS \cite{Gromacs} was used to
perform all the atomistic (AA and UA) MD simulations. Initial
well-equilibrated atomistic polymer melts are obtained by back-mapping CG
melts \cite{vagelis2}. All simulations have been performed at
constant temperature and volume ($NVT$ ensemble) at
experimental densities \cite{PS_dens}. The overall simulation time
of the production runs ranged from 50 ns to 300 ns depending on
the molecular weight.

\begin{figure}
\begin{center}
\includegraphics[scale=0.4]{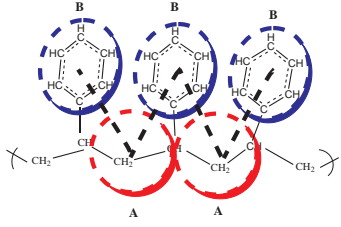}
\caption{Coarse-graining mapping scheme of PS: one monomer is
mapped to two different CG beads
($\sigma_{A}=$4.1\AA$ $, $m_{A}$=27amu and
$\sigma_{B}=$5.2\AA$ $, $m_{B}$=77amu)
\cite{vagelis2}.}
\label{fig:ps_cg_mapping}
\end{center}
\end{figure}

CG MD simulations have been performed using a CG model for PS in
which one PS monomer is mapped onto two effective coarse grained
beads, i.e. a 2:1 model \cite{vagelis2} (see Figure
\ref{fig:ps_cg_mapping}). 
CG bead "A" corresponds to the CH$_{2}$ of a PS monomer plus the half mass of
each of the two neighboring CH groups along the chain backbone,
whereas CG bead "B" is the phenyl ring.
This model does not loose too many
structural details in comparison to all atom systems, while still
being very efficient compared to atomistic simulations. Various
monodisperse atactic PS melts, with molecular weight from 1kDa up
to 50kDa have been studied with CG MD simulations.
Note that the characteristic molecular weight $M_e$ for the
formation of entanglements for atactic PS is about 15kDa (at
$T$=463$K$) \cite{lodge1}. All CG systems have been originally
generated by a combination of Monte Carlo and
MD simulations following the procedure by Auhl et al.
\cite{auhl,vagelis2}.
CG MD simulations are performed in dimensionless LJ units using $m_A$ to
scale all masses, $\sigma_{AV}=(\sigma_{A}+\sigma_{B})/2$ to scale
all lengths and $\epsilon=kT$ to scale all energies. In order to
control the temperature in the system we use a Langevin thermostat
with friction coefficient $\Gamma=1.0\tau^{-1}$. All
coarse-grained MD simulations are performed using the ESPResSO
package \cite{espresso} with a time step \emph{$\Delta$t}=0.01
$\tau$, with $\tau=\sqrt{m_A\sigma_{AV}^2/\epsilon}$. MD
simulations ran for times  between $1 \times 10^{4} \tau$ and $ 3
\times 10^{6} \ \tau$ depending upon the system size.

\textit{Time Mapping.}$-$
The problem of time scales can be discussed in terms of local
(monomeric) friction coefficients.
For both the Rouse as well as the reptation model the local motion
is governed by a scalar friction coefficient $\zeta$, so that the
melt viscosity $\eta \propto \zeta$ and the chain diffusion
constant $D(N) \propto \zeta^{-1}$. For the modeling of the
polymer melt this bead friction depends on the specific
representation of the polymer. The softer CG
potentials result in a significantly reduced
effective friction coefficient, $\zeta^{CG}$, between beads
compared to the friction coefficient in the united atom
description, $\zeta^{UA}$, and the all atom description,
$\zeta^{AA}$, respectively. As a consequence the time in the
dynamic CG simulations does not correspond to the real time of the
polymeric system and has to be properly scaled. In most cases the
differences between all atom and united atom simulations are much
smaller then the typical error bars of a simulation. For the
TraPPE-UA PS model however this is not the case. Though in general a
disadvantage, for PS this makes it possible to obtain a proper
scaling by employing the AA-UA-CG hierarchy. Thus the time
scaling parameter, which is the ratio of the effective bead
frictions in the atomistic and the CG description, is given by
\begin{equation}
  \label{eq:scale}
  S(M) \equiv \zeta^{AA}/\zeta^{CG}= S_{AA-UA}(M)\cdot S_{UA-CG}(M)
\end{equation}
Because the local energy landscape is quite complex and
fluctuating it is not possible to give a well founded analytical
prediction of $S$. Therefore, $S$ should be obtained using data
taken either from atomistic simulations, as done here, or directly
from experiment.

\begin{figure}
\begin{center}
\includegraphics[scale=0.25]{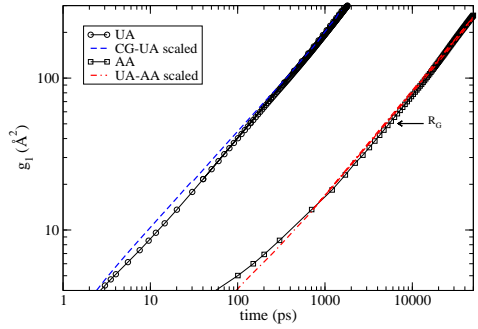}
\caption{Time scaling of the CG simulations using UA data and of
the UA simulations using atomistic data for a PS melt ($M$=1kDa,
$T$=463$K$), based on the motion of the polymer beads.}
\label{fig:timemapping_seg}
\end{center}
\end{figure}

To determine $S$ we match mean square displacements (MSD) of chain
beads over a considerable time, where both amplitude and shape
coincide. First we examine the time mapping of the CG data based
on the UA MD simulations. Figure \ref{fig:timemapping_seg} shows
the mean square displacements averaged over all beads $i$ of the
CG model and the correspondingly analyzed UA model (circles), $g_1(t)$,
$g_1(t)= \{\langle(r_i(t)-r_i(0))^2\rangle\}_{\langle i \rangle}$
from UA MD and CG MD simulations for a specific PS melt
(1kDa, T=463K). The scaling factor, $S_{UA-CG}$, in order to match
the two curves on top of each other in the long time regime, gives
$S_{UA-CG}$(\textit{M}=1kDa)= 3.1ps/$\tau$. With this scaling
factor both curves coincide above a distance of about 10 \AA$ $
and a corresponding time of about 200 ps. Below that distance and
time the coarse graining results in a different shape of the
curve, illustrating that a mere crossing of the curves is not
sufficient to determine $S$.
Note also that $S_{UA-CG}$
is the same if the mapping is based on a quantity describing
the orientational dynamics (for example the end-to-end vector
autocorrelation function) of the PS chains \cite{vagelis_long}.

\begin{figure}
\begin{center}
\includegraphics[scale=0.25]{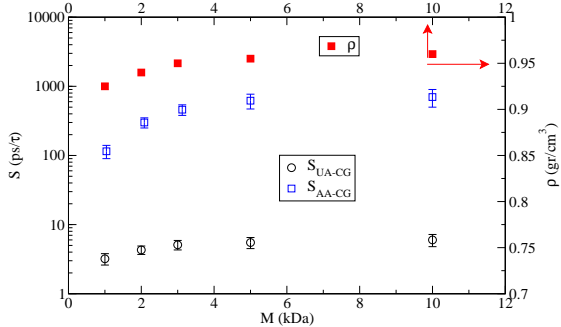}
\caption{Time mapping of the CG simulations of the PS melts using
UA and AA data, and density as a function of $M$ ($T$=463K).}
\label{fig:timemapping_MW}
\end{center}
\end{figure}

In general $S_{UA-CG}$ depends on molecular weight and density,
which depends on $M$ (chain end free volume effect), as shown in
Figure \ref{fig:timemapping_MW} (circles) for the systems studied
by both UA MD and CG MD simulations. As we can see $S_{UA-CG}$
varies in the short length regime (up to about 50 monomers),
ranging from 3.1ps/$\tau$ to about 6.0ps/$\tau$, and then it
remains constant. This is in phase with the observed change in
density, which varies from 0.925g/cm$^{3}$ for 1kDa to about
0.97g/cm$^{3}$ for the 10kDa and higher molecular weight melts
(see Figure \ref{fig:timemapping_seg}) \cite{PS_dens}.
At high molecular weights (above 10kDa) the change
in the polymer dynamics is entirely due to the increase of the
molecular weight.
On the other hand, in the short length regime
the density effect is very important.
The latter one is not
being described accurately in the CG model, resulting into a
dependence of $S$ on the density (and on the molecular length).

The important result of Figure \ref{fig:timemapping_MW}, is that a
single value for the time scaling parameter $S$ is appropriate to
describe the dynamics of long polymer chains. However the UA model
itself includes some minor coarse graining and in the case of the
TraPPE-UA PS model predicts faster dynamics than the all-atom simulations.
Thus we have to
follow a same procedure as above, however now for the two models
exhibiting atomistic detail. A typical example is shown in Figure
\ref{fig:timemapping_seg} (squares). Though qualitatively similar,
there is a remarkable quantitative difference. As Figure
\ref{fig:timemapping_seg} displays the two sets of data perfectly
match from a distance above about 3\AA$ $, this is for
significantly smaller length scales than before, of the order of
the size of a benzene ring ($R_{G}$ is shown with arrow).
For the present case of Figure
\ref{fig:timemapping_seg} we arrive at $S_{AA-UA}$(\textit{M}=1kDa)=35,
resulting in $S_{AA-CG}$(\textit{M}=1kDa)=110ps/$\tau$. This together with
computational efficiency of the TraPPE UA model allows us to
determine the time scaling factor with its full chain length
dependence (up to \textit{M}=10kDa), as shown in Figure
\ref{fig:timemapping_MW} (squares). Note the similar qualitative
but the large quantitative difference between $S_{UA-CG}(M)$ and
$S_{AA-CG}(M)$. Alternatively one can follow the observation, that
the variation of $S$ follows approximately the changes in density
rather than the molecular weight itself, even though this density
change is due to the chain length variation.
This is not surprising since, for polymers both density and
monomeric friction coefficient vary with molecular weight
because of the higher chain end free volume (free volume theory)
and two melts with the same density would be expected to have the same
friction coefficient \cite{ferrybook}.
Therefore by performing the time mapping for the short chain system but at the
density of the longer chains one also can obtain a
reliable estimate of $S_{AA-UA}$. If we follow this procedure the
combined time mapping
$S_{AA-CG}(M)$
varies between $\simeq$ 110ps/$\tau$ (for the 1kDa
system) and $\simeq$ 700ps/$\tau$ for the high (10kDa and above)
molecular weight (polymeric) regime. Here we follow the latter
because of the extensive computational effort needed to obtain
reliable data for MSDs of long all-atom PS chains.
$S_{AA-CG}$ now can be applied to determine absolute
diffusion constants in long chain PS melts, based on CG
simulations.

\begin{figure}
\begin{center}
\includegraphics[scale=0.3]{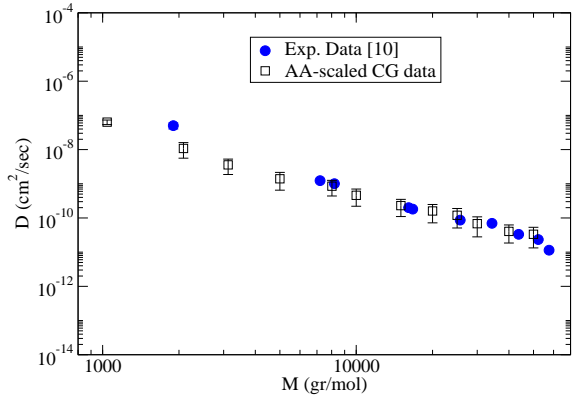}
\caption{Self-diffusion coefficient of PS melts as a function
of the molecular weight ($T$=463$K$).}
\label{fig:dvsn_sc}
\end{center}
\end{figure}

\textit{Self-Diffusion Coefficient of Polystyrene Melts.}$-$
Following the hierarchical methodology described above, the
dynamical and rheological properties of long entangled polymeric
systems can be obtained from the CG dynamic simulations in real
units. Here we focus on the self-diffusion coefficient, $D$, which
can be calculated directly from the MSD of the chain
center-of-mass by rescaling 
$\tau$ to sec by the above described procedure.
Data for the self-diffusion
coefficient of the CG PS melts, scaled with
$S_{AA-CG}(M)$, as a function of the
molecular weight (squares) are shown in Figure \ref{fig:dvsn_sc} and are
compared to experimental data (circles) \cite{lodge1}.
Both simulation and experimental data are not corrected for
the chain end free volume.
The range of molecular weights (up to 50kDa) spans the regime from
unentangled to entangled PS melts.

The results show a remarkable qualitative and quantitative
agreement between the experimental and the simulated diffusion
coefficients (see figure \ref{fig:dvsn_sc}).
This is of
particular importance if we consider that results from the CG
dynamic simulations are compared to experimental data, by using
only detailed atomistic simulations for a few reference
short-chain systems, \emph{without any adjustable parameter}. The
error bars in the AA-scaled CG data are mainly due to the high
error bars in the underlying all-atom data. The larger deviation
between the simulation and the experimental data in the short
length regime is not surprising if we consider the effect of the
(small) polydispersity of the experimental data ($I \simeq$ 1.04):
in these short chains the presence of even only a small amount of
PS oligomers acts like an effective dilution and can increase the
free volume and the diffusion of the systems.

Note also that the entire regime of both simulation and
experimental data can be fitted using a power-law dependence ($D
\sim M^{-b}$)
with a single exponent of about $b\approx 2.1 \pm 0.2$.
For short chain polymer melts the
Rouse model predicts $b$=1 while long entangled polymer melts the
pure reptation theory predicts $b$=2 \cite{doiedw}. However the Rouse
model neglects the molecular weight dependence of the density and
and thus of the friction coefficient. In order to eliminate this
effect a correction can be made for $D$ as it is often done in the
analysis of experiments; this will be a part of a future work
\cite{vagelis_long}.


It is of interest to estimate the time scales involved in the
dynamics of the PS systems studied here: the diffusion coefficient
of the higher molecular weight PS melts (50kDa) is of the order of
about 10$^{-11}$cm$^2$/sec. This results into a relaxation time of
the whole chain - decorrelation of the end-to-end vector -,
$\tau_d$ (according to reptation theory $\tau_d=<R^2>/3\pi^2D$
\cite{doiedw}) of about 6.0ms, many orders of magnitude longer
that what can be modeled with atomistic molecular dynamic
simulations in such systems.

In conclusion we present a hierarchical simulation approach that
combines dynamic simulations on different length and time
scales. A mapping over a small range of molecular lengths, using
atomistic and united-atom simulation data, shows that the time
mapping parameter $S$ varies as a function of chain length induced
by different melt densities. The asymptotic plateau value of $S$ can
be used for scaling the CG dynamic results of longer polymeric
chains, where it is not possible to have reliable atomistic data
at all.

Furthermore the multi-scale nature of the proposed scheme,
combined with a proper back-mapping procedure
\cite{vagelis2,actapaper1,peter}, allows to
study time scales \textit{ranging from a few fs up to ms} by MD
simulations and to compare directly to experimental data without
any adjustable parameter. The accuracy of the CG dynamical
predictions, compared to experimental data, of course depends on
the accuracy of the underlying microscopic (atomistic) model with
which the CG ``raw'' data are scaled. This opens up the way for
developing simulation methodologies that can be used for
quantitative studies of the dynamics and the rheology of complex
systems within the $M$ regime relevant to the polymer processing.
The proposed approach can be directly extended to other polymer chemistries
by properly choosing a structure-based CG model, which reproduces
structural properties for length scales of about 0.5-1nm and larger.
Furthermore it can be used for the study such different
systems as non-equilibrium polymer melts, polymers at temperatures
near to T$_{g}$, or polymer/solid interfacial systems.

One the other hand, and despite the success of the proposed methodology,
still a number of issues remain. For example, for
multi-component systems or polymers in solution, the nature of the
friction might not allow a renormalization in time with a single
quantity. For such systems molecular dynamics simulations could be
used for the calculation of the entire friction matrix, which can
be then involved in the equations of motion on an even more
coarse level, as for example in the level of the primitive paths
\cite{kk_science} or through the GENERIC formalism \cite{ottinger1}.

\textit{Acknowledgments.}$-$ Very fruitful discussions with Nico van der Vegt,
Dirk Reith and Burkhard D\"unweg are greatly appreciated.


\clearpage


\newpage
\begin{thebibliography}{99}

\bibitem{doiedw} M. Doi and S.F. Edwards,
\emph{The Theory of Polymer Dynamics} (Claredon Press, Oxford, England, 1986).

\bibitem{ferrybook} J.D. Ferry, \emph{Viscoelastic Properties
of Polymers} (John Wiley and Sons, New York, 1980).

\bibitem{binderbook} \emph{Monte Carlo and
Molecular Dynamics Simulations in Polymer Science} edited by K.
Binder, (Oxford University Press, New York, 1995).

\bibitem{mcleish1} T. Mcleish,
Adv. Phys. \textbf{51}, 1379 (2002).

\bibitem{kk_erice} K. Kremer, in \emph{Proceedings of the
International School of Solid State Physics - 34$^{th}$ Course:
Computer Simulations in Condensed Matter: from Materials to
Chemical Biology} ed. by K. Binder, K. and G. Ciccoti (Erice,
2006).

\bibitem{harmanda1} V.A. Harmandaris and V. Mavrantzas,
Molecular Dynamic Simulations of Polymers in
\emph{Simulation Methods for Polymers}, ed. by D.N. Theodorou and
M. Kotelyanski (Marcel Dekker, 2004); V. Harmandaris \textit{et al.},
Macromolecules, \textbf{36}, 1376 (2003).


\bibitem{baschnagel} J. Baschnagel \textit{et al.},
Adv. Polym. Sc. \textbf{152}, 41 (2000).


\bibitem{actapaper1} W. Tsch\"op \textit{et al.}, Acta Polym. \textbf{49}, 61, (1998);
S.O. Nielsen, G. Srinivas, and M. Klein, J. Chem. Phys. \textbf{123}, 124907 (2005);
V.A. Harmandaris \textit{et al.}, Macromolecules \textbf{39}, 6708 (2006);
B. Hess \textit{et al.}, Soft Matter \textbf{2}, 409, (2006);
J. Padding and W.J. Briels, J. Chem. Phys. \textbf{117}, 925 (2002);
W. Paul and N. Pistoor, Macromolecules \textbf{27}, 1249 (1994).



\bibitem{vagelis2} V.A. Harmandaris \textit{et al.},
Macrom. Chem. and Phys. \textbf{208}, 2109 (2007);
V.A. Harmandaris  \textit{et al.}, Macromolecules \textbf{40}, 7026 (2007).









\bibitem{lodge1} T.P. Lodge, Phys. Rev. Lett. \textbf{83}, 3218 (1999);
M. Antonietti, K.J. F\"olsch, and H. Sillescu, Makromol. Chem. \textbf{188}, 2317 (1987);
O. Urakawa \textit{et al.}, Macromolecules \textbf{37}, 1558 (2004);
C. Liu, J., He, E. van Ruymbeke, R. Keunings, and C. Baily,
Polymer \textbf{47}, 4461 (2006).





\bibitem{plathe2} F. M\"uller-Plathe
Macromolecules \textbf{29}, 4782 (1996).

\bibitem{trappe1} C.D. Wick, M.G. Martin, and J.I. Siepmann,
J. Phys. Chem. B \textbf{104}, 8008 (2000).

\bibitem{Gromacs} H.J.C. Berendsen, D. van der Spoel, and R. van
Drunen, Comp. Phys. Comm. \textbf{91}, 43 (1995).

\bibitem{PS_dens} P. Zoller and D.J. Walsh,
\emph{Standard Pressure-Volume-Temperature Data for Polymers} (Technomic, Lancaster, 1995).

\bibitem{auhl} R. Auhl \textit{et al.},
J. Chem. Phys. \textbf{119}, 12718 (2003).

\bibitem{espresso} A. Arnold \textit{et al.},
Comp. Phys. Comm \textbf{174}, 704 (2006).

\bibitem{vagelis_long}
V.A. Harmandaris and K. Kremer, in preparation.

\bibitem{peter} C. Peter, L. Delle Site, and K. Kremer,
Soft Matter \textbf{4}, 859 (2008).

\bibitem{kk_science} R. Evereaers \textit{et al.},
Science \textbf{303}, 823 (2004); S.K. Sukumaran \textit{et al.},
J. Polym. Sci. Polym. Phys. \textbf{43}, 917 (2005).


\bibitem{ottinger1} H. \"Ottinger, \emph{Beyond Equilibrium Thermodynamics}
(John wiley $\&$ Sons, New Jersey, 2005).

\end{thebibliography}
\end {document}